\documentclass[showpacs,twocolumn]{revtex4}
\usepackage{amsmath}
\usepackage{graphicx}

\newcommand{\calM}{{\cal M}}
\newcommand{\calO}{{\cal O}}
\newcommand{\calC}{{\cal C}}

\begin{document}
\title{Narrow-width approximation limitations}
\date{\today}
\author{N.~Kauer}
\affiliation{Institut f\"ur Theoretische Physik, Universit\"{a}t W\"{u}rzburg, D-97074 W\"{u}rzburg, Germany}
\begin{abstract}
The quality of the narrow-width approximation is examined for partonic and 
convolved cross sections of sample processes.  By comparison with accurate 
predictions significant limitations are revealed.
\end{abstract}
%25.75.Dw Particle and resonance production
%11.80.Fv Approximations (eikonal approximation, variational principles, etc.)
%12.60.-i Models beyond the standard model
\maketitle

The narrow-width approximation (NWA) is widely applied to predict the probability for 
resonant scattering processes when the total decay width $\Gamma$ of the resonant 
particle is much smaller than its mass $M$. 
In the limit $\Gamma\to 0$, the squared propagator 
$ D(q^2)\equiv \left[\left(q^2-M^2\right)^2+(M\Gamma)^2\right]^{-1}$ with 4-momentum $q$
approaches $ D(q^2) \sim K\delta(q^2-M^2)$ with 
$K=\pi/(M\Gamma)=\int_{-\infty}^\infty dq^2  D(q^2)$.  
The scattering cross section $\sigma$ thus approximately decouples into on-shell 
production ($\sigma_p$)
and decay as shown in Eq.~(\ref{eq:nwa}) for a scalar process \cite{footnote1}.  
The generalization to multiple resonances is straightforward.
We use the notation of \cite{Yao:2006px}, Sec.~38.
Based on the scales occurring in $ D(q^2)$, the conventional error 
estimate is $\calO(\Gamma/M)$.
The branching fraction for the considered decay mode is then given by
$\text{BR} = \Gamma_X/\Gamma = \sigma_\text{NWA}/\sigma_p$, where
$\Gamma_X$ is the partial decay width.  We define the effective
branching ratio $\text{BR}_\text{eff} \equiv \sigma/\sigma_p$, which is
related to the branching ratio via $\text{BR}_\text{eff} = (1+R)\,\text{BR}$
with $R \equiv \sigma/\sigma_\text{NWA}-1= R^{(1)}+{\cal O}(\Gamma^2)$.

\begin{figure}
\includegraphics[height=2.8cm]{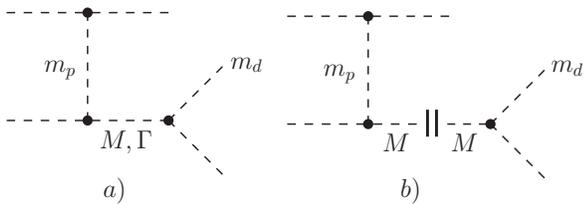}
\caption{\label{fig:process1}Scalar process with (a) Breit-Wigner propagator and (b) in NWA. The double bars indicate the decoupling of production and decay. Lines without labels correspond to massless particles.}
\end{figure}

\begin{figure}
\includegraphics[width=5.5cm]{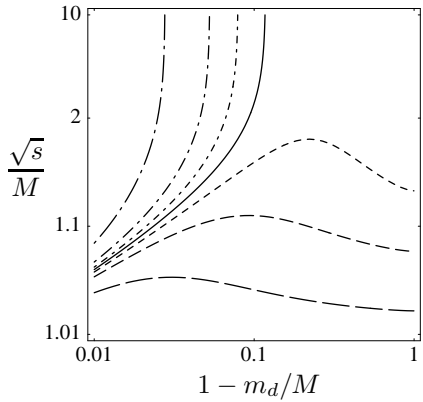}
\caption{\label{fig:rtilde_process1_contour}
Contour plot of $R\equiv\sigma/\sigma_\text{NWA}-1$ normalized to $\Gamma/M=0.01$ 
as function of $\sqrt{s}$ and $m_d$ for the process of Fig.~\protect\ref{fig:process1} 
with $m_p=0.02\,M$.
Contour lines are shown for $R=0$ (solid), $R>0$ (dot-dashed) and $R<0$ (dashed)
with $|R|/(\Gamma/M)\in \{1,3,10\}$.  The dash length increases with magnitude.
}
\end{figure}

To examine the conventional error estimate, we calculate $R^{(1)}$ for the
scalar process of Fig.~\ref{fig:process1}.
Here, $|\calM_p|^2\propto 1/(t-m_p^2)^2$ %
\cite{footnote2}
and $|\calM_d|^2\propto 1$.
\begin{widetext}
\begin{equation}
\label{eq:nwa}
\sigma = \frac{(2\pi)^7}{2s}\int_{q^2_\text{min}}^{q^2_\text{max}}dq^2\int d\phi_p|\calM_p(q^2)|^2 D(q^2)\int d\phi_d|\calM_d(q^2)|^2
\;\approx\;\sigma_\text{NWA}\equiv\frac{(2\pi)^7}{2s}\int d\phi_p|\calM_p(M^2)|^2 K \int d\phi_d|\calM_d(M^2)|^2
\end{equation}
\begin{gather}
R^{(1)} = \left\{
\frac{M \left(s-m_d^2\right)}{\pi \left(m_d^2-M^2\right)  \left(s-M^2\right)}
+
\left[\pi M \left(m_d^2-M^2\right) \left(s-M^2\right) \left(s+m_p^2\right)
\left(s-M^2+m_p^2\right)\right]^{-1}
\Bigg[m_d^2\,s\left(s-M^2+m_p^2\right)^2 \ln\frac{s}{m_d^2}
\right.\nonumber\\
\left.\left.
+\left(s+m_p^2\right)
\bigg(m_d^2 \Big(s \big(s+m_p^2\big)-2 M^2 s+M^4\Big)-M^4 m_p^2\bigg) \ln\frac{M^2-m_d^2}{s-M^2}
+M^4 m_p^2
\left(s-m_d^2+m_p^2\right) \ln\frac{s-m_d^2+m_p^2}{m_p^2}\right]
\right\}\cdot\Gamma
\label{eq:r1scalar}
\end{gather}
\end{widetext}
The result in Eq.~(\ref{eq:r1scalar})
shows that $R$ and hence $\text{BR}_\text{eff}$ depend not only on 
$\Gamma$ and $M$, but on additional scales introduced through the center of mass 
energy $\sqrt{s}$ or particle masses in production or decay.  Selection cuts have 
a similar impact that is not explicitly considered here.  Note that the actual 
relative deviation in units of the conventionally expected error, i.e.~$R/(\Gamma/M)$, 
is width independent to first approximation in $\Gamma$.
As the threshold $\sqrt{s}=M$ is approached the on-shell production 
phase space volume vanishes and $\sigma_\text{NWA}\to 0$ while $\sigma$ remains finite.
$R$, hence, diverges and the NWA becomes unreliable.
From Eq.~(\ref{eq:r1scalar}) one also sees that $R$ diverges
in the limit $m_d\to M$, i.e.~when the sum of the daughter masses approaches 
the mass of the decaying particle.
The NWA can thus not be used for all decay modes in scenarios with almost degenerate 
mass hierarchies.
We illustrate the behaviour of $R$ in Fig.~\ref{fig:rtilde_process1_contour}.  
For $m_d\lesssim0.9\,M$ and $\sqrt{s}\gtrsim1.1\,M$,
the NWA uncertainty is indeed of $\calO(\Gamma/M)$.
Note that while $R/(\Gamma/M)\lesssim -10$ for $\sqrt{s}\gtrsim1.01\,M$, it turns
positive for yet smaller values of $\sqrt{s}$
%, e.g. at $\sqrt{s}\approx 1.0001\,M$ for $m_d=0.5\,M$, $\Gamma=0.01\,M$ and $m_p=0.02\,M$
(see also Fig.~\ref{fig:rtilde_process1_0.9}).
The $m_p$-dependence of the ratio $R$ is in general relatively weak 
despite the mass singularity in the production cross section.
We note that significant deviations in the threshold region have been 
observed in connection with Standard Model (SM) predictions 
for $W$-pair production and decay in $e^+e^-$ collisions \cite{Wpred}.

\begin{figure}
\includegraphics[height=4.5cm]{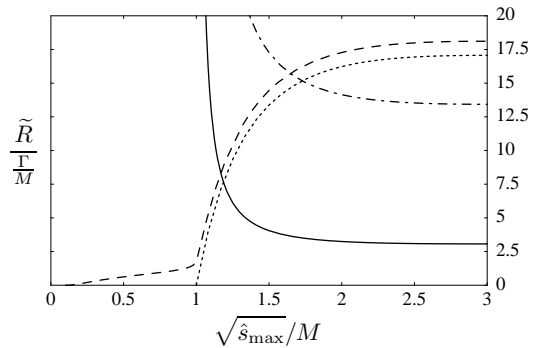}
\caption{\label{fig:rtilde_process1_0.1_0.001}
The cross sections $\widetilde\sigma$ (dashed) and $\widetilde\sigma_\text{NWA}$ 
(dotted) in unspecified normalization and 
$\widetilde R\equiv\widetilde\sigma/\widetilde\sigma_\text{NWA}-1$ 
normalized to $\Gamma/M$ (solid) are shown
as functions of $\sqrt{\hat{s}_\text{max}}$ (see main text) 
for the process of Fig.~\protect\ref{fig:process1} 
with $\Gamma=0.02\,M$, $m_p=0.01\,M$, $m_d=0.1\,M$ and $\sqrt{s}=3\,M$.
$\widetilde R/(\Gamma/M)$ with $m_d=0.001\,M$ is also displayed (dot-dashed).
}
\end{figure}

\begin{figure}
\includegraphics[height=4.5cm]{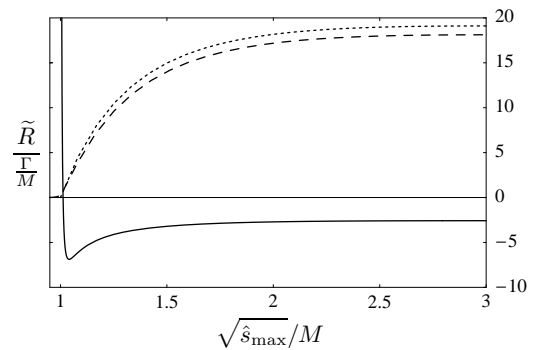}
\caption{\label{fig:rtilde_process1_0.9}
The cross sections $\widetilde\sigma$ and $\widetilde\sigma_\text{NWA}$ 
as well as $\widetilde R$ are shown for the process of Fig.~\protect\ref{fig:process1}
with $m_d=0.9\,M$. Other details as in Fig.~\protect\ref{fig:rtilde_process1_0.1_0.001}.
}
\end{figure}

Resonant cross sections typically peak just above threshold.  
In Figs.~\ref{fig:rtilde_process1_0.1_0.001} and \ref{fig:rtilde_process1_0.9}, 
we illustrate the accuracy of the NWA above 
threshold for reactions with distributed $\sqrt{s}$ by
displaying $\widetilde\sigma_{[\text{NWA}]}(\sqrt{\hat{s}_\text{max}})\propto 
1/s\int_0^{\hat{s}\,\text{max}}
d\hat{s}\,\sigma_{[\text{NWA}]}(\hat{s})\ln(s/\hat{s})$ corresponding to constant 
incoming particle (parton) momentum 
distribution functions (PDFs) and $\widetilde R\equiv\widetilde\sigma/\widetilde\sigma_\text{NWA}-1$ normalized to $\Gamma/M$.  Like $R/(\Gamma/M)$, $\widetilde R/(\Gamma/M)$
does not depend on $\Gamma/M$ in first approximation.
At hadron colliders, the convolution with steeply falling PDFs enhances
the weight of the region around and below threshold.  It also affects the 
resonance line shape, i.e.~alters the $d\sigma/dq^2$ distribution, whereas
in NWA the Breit-Wigner is integrated out unmodified into $K$.

\begin{figure}
\includegraphics[height=2.8cm]{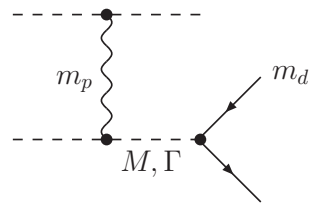}
\caption{\label{fig:process2}Process with non-scalar particles in production 
(gauge vector boson) and decay (fermions).}
\end{figure}

\begin{figure}
\includegraphics[width=5.5cm]{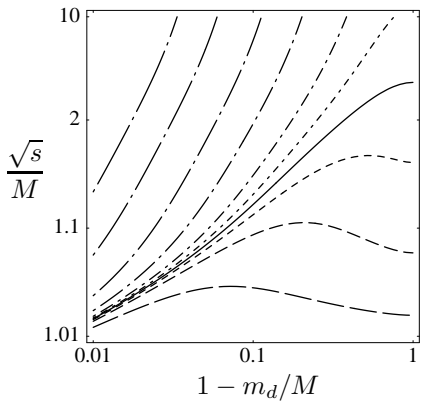}
\caption{\label{fig:rtilde_process2_contour}
Contour plot of $R$ as function of $\sqrt{s}$ and $m_d$ for the process of 
Fig.~\protect\ref{fig:process2}.
Contour lines for $R/(\Gamma/M)\in \{-10,-3,-1,0,1,3,10,30,100,300\}$ are shown.
Other details as in Fig.~\protect\ref{fig:rtilde_process1_contour}.
}
\end{figure}

\begin{figure}
\includegraphics[height=4.5cm]{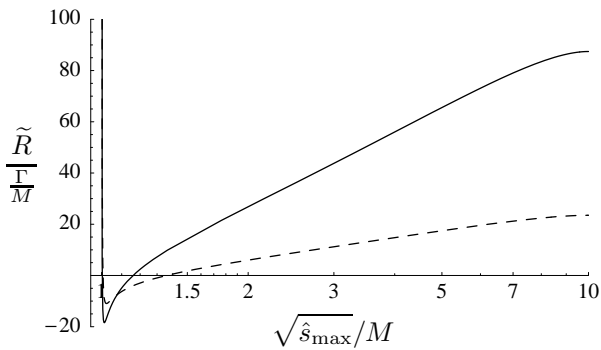}
\caption{\label{fig:rtilde_process2_0.9_0.95}
$\widetilde R$ is shown as function of $\sqrt{\hat{s}_\text{max}}$ 
for the process of Fig.~\protect\ref{fig:process2} with $m_d=0.95\,M$ (solid) and
$m_d=0.9\,M$ (dashed) for $\sqrt{s}=10\,M$.
Other details as in Fig.~\protect\ref{fig:rtilde_process1_0.1_0.001}.
}
\end{figure}

\begin{figure}[htb]
\includegraphics[height=4.5cm]{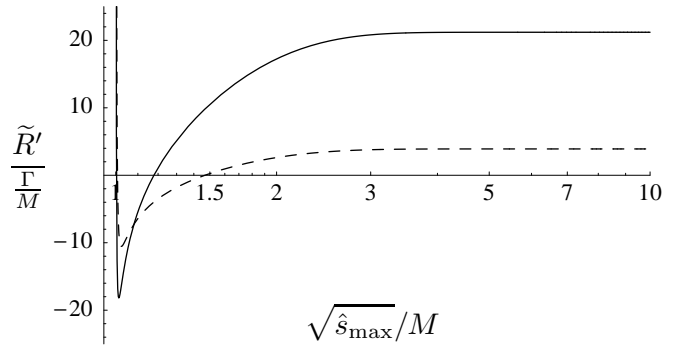}
\caption{\label{fig:rtildeprime_process2_0.9_0.95}
$\widetilde R'\equiv\widetilde\sigma'/\widetilde\sigma'_\text{NWA}-1$
with cross sections $\widetilde\sigma'$ and $\widetilde\sigma'_\text{NWA}$ 
that involve steeply falling PDFs (see main text)
is shown as function of $\sqrt{\hat{s}_\text{max}}$ 
for the process of Fig.~\protect\ref{fig:process2}
with $m_d=0.95\,M$ (solid) and
$m_d=0.9\,M$ (dashed) for $\sqrt{s}=10\,M$.
Other details as in Fig.~\protect\ref{fig:rtilde_process1_0.1_0.001}.
}
\end{figure}

Next, we consider a process with non-scalar particles
and more complex interactions.  More specifically, we repeat our analysis for the 
process in Fig.~\ref{fig:process2}. Here, 
$|\calM_p|^2\propto (t + 2 s - q^2)^2/(t-m_p^2)^2$ and $\sum_\text{spin}|\calM_d|^2\propto q^2-m_d^2$.
For this process, the $\calO(\Gamma)$-result for $R$ corresponding to Eq.~(\ref{eq:r1scalar}) contains dilogarithms and is too complex to be displayed here.
Fig.~\ref{fig:rtilde_process2_contour} shows that for this process
comparable enhancements occur already at less extreme parameter values.
As shown in Fig.~\ref{fig:rtilde_process2_0.9_0.95}, the large off-shell increase
for higher $\sqrt{s}$ leads to a continued rise of $\widetilde R$
for constant PDFs.  
With the steeply falling PDFs characteristic for hadron colliders, 
however, the ratio saturates as seen in 
Fig.~\ref{fig:rtildeprime_process2_0.9_0.95} for 
$\widetilde\sigma'_{[\text{NWA}]}(\sqrt{\hat{s}_\text{max}})\propto$
$1/s\int_0^{\hat{s}\,\text{max}}
d\hat{s}\,\sigma_{[\text{NWA}]}(\hat{s}) \int_{\hat{s}/s}^1dx/x\,f(x)f(\hat{s}/(xs))$
with $f(x)\propto (1-x)^9/\sqrt{x}$.
\begin{figure}
\includegraphics[height=2.8cm]{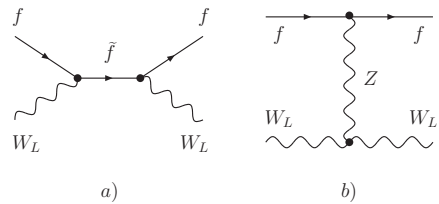}
\caption{\label{fig:process3} Scattering of a massless, neutral fermion and a 
longitudinally polarized $W^-$ boson in the GWS model with (a) $s$-channel ($\calM_s$) 
and (b) $t$-channel ($\calM_t$) amplitude contributions.  For 
$\sqrt{s}\approx m_{\widetilde f}\equiv M > m_W\equiv m$ the $s$-channel contribution 
will be resonant ($\Gamma_{\widetilde f}\equiv \Gamma$). The $t$-channel contribution
with $m_Z=m/\cos\theta_W$ is nonresonant.}
\end{figure}
In addition to resonant amplitudes, typically nonresonant or subresonant amplitudes 
contribute, too, but are expected to be suppressed.  We discuss the effect of these 
contributions by means of the scattering of a massless fermion with electric charge 
$Q_f=0$ and a longitudinally polarized $W^-$ boson in the GWS model 
(see \cite{GWSmodel} and \cite{Yao:2006px}, Sec.~10).  The contributing Feynman
graphs are shown (and our notation is clarified) in Fig.~\ref{fig:process3}.
The corresponding amplitudes are given by $\calM_s=\calC\cdot s/m^2/(s-M^2+iM\Gamma)$
and $\calM_t=\calC\cdot (2-(s-3m^2)(s+m^2)t/m^2/(s-m^2)^2)/(t-m_Z^2)$ with
$\calC = -ig^2/2\sqrt{(s-m^2)^2+s\, t}$, where $g$ is the weak coupling constant.
Close to resonance, the $\calM_s$ contribution is
expected to exceed the $\calM_t$ contribution by $\calO((M/\Gamma)^2)$.
In our example with a neutral external fermion this is indeed the case.
For $|Q_f|\neq 0$, however, the additional graph with
$t$-channel exchange of a massless photon will make the nonresonant
contribution divergent in the zero-mass limit for the external fermion,
while the resonant and $Z$-exchange contributions do not diverge
for finite $m$.  Nonresonant contributions can thus be larger than anticipated.
After integrating the Breit-Wigner 
a resonant enhancement of the cross section by $\calO(M/\Gamma)$ is expected. 
This requires that the contribution from resonant graphs has to dominate 
not only close to resonance, but also in the region with $|\sqrt{s}- M|\gtrsim\Gamma$, 
which contributes 30\% even when 
nonresonant graphs are negligible.
If the affected phase space regions can be distinguished 
and eliminated experimentally, these effects can be suppressed.
We note that a significant impact of sub- and nonresonant contributions has been 
observed for $W$ boson \cite{W_nonres} and top quark \cite{top_nonres} production 
and decay in the SM.

\begin{figure}
\includegraphics[width=7.8cm]{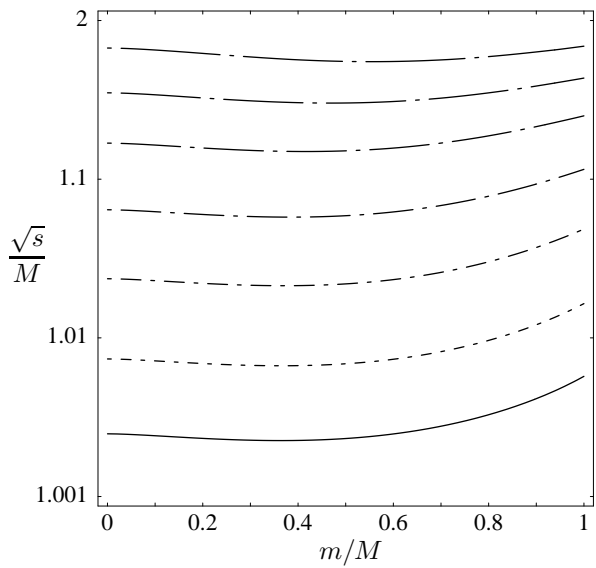}
\caption{\label{fig:r_interference_process3_contour}
Contour plot of $R_i\equiv(\sigma(\calM_s)+\sigma(\calM_t))/\sigma(\calM_s+\calM_t)-1$ normalized to 
$\Gamma/M=0.01$ with the resonant $s$- and nonresonant $t$-channel cross section contributions 
for the process of Fig.~\protect\ref{fig:process3}
as function of the invariant mass of the resonant particle $\sqrt{s}$ and 
$W$ mass $m$ for $\cos\theta_W=0.8$.
Contour lines are shown for 
$R_i/(\Gamma/M)=1$ (solid) and $R_i/(\Gamma/M)\in \{3,10,30,100,300,1000\}$ (dot-dashed).  The dash length increases with magnitude.
}
\end{figure}
For processes with significant nonresonant amplitude ($\calM_{nr}$) contribution
one could apply the NWA to the resonant amplitude ($\calM_r$) contribution and 
separately calculate and analyze the $\calM_{nr}$ contribution after one has 
established that the interference between both is negligible \cite{footnote3}.
Fig.~\ref{fig:r_interference_process3_contour} exemplifies that 
$\calM_r$-$\calM_{nr}$ interference effects can be larger than 
the expected NWA uncertainty, even close to resonance.
Note that the interference contribution can 
not be calculated in NWA, because then $\calM_r$ would have to be integrated with the 
on-, but $\calM_{nr}$ with the off-shell phase space.  
More accurate calculations that do not rely on the NWA are therefore always 
necessary to assess interference effects.  

Having established various limitations of the NWA by considering selected examples, we
conclude that 
more detailed studies for phenomenologically relevant processes are necessary
to guarantee accurate predictions.
We have investigated the quality of the 
NWA in the Minimal Supersymmetric Standard Model
and find inaccuracies for relevant processes in not-ruled-out scenarios 
that are an order of magnitude larger than the conventional uncertainty 
estimate $\calO(\Gamma/M)$ of typically 1-10\% \cite{Berdine:2007}.
This study also shows that secondary decays in cascades do in general 
not mitigate the deviation and in selected scenarios enhance it even further.

Beyond leading order, at least QCD corrections should be taken into account since 
they are generally of similar size and stabilize the scale dependence.
The 1-loop multi-leg techniques required for a combined analysis
are becoming available (see e.g.~\cite{Buttar:2006zd}, Sec.~30, and refs. therein).
Our analysis could be extended to the at NLO level often used on-shell pole
approximation \cite{poleapprox}, e.g. by comparison with results obtained with
the complex-mass scheme \cite{complexmassscheme}.
The presented results further indicate that scale ratios other than $\Gamma/M$ may have to
be taken into account when effective-theory-inspired methods are applied 
\cite{efftheor}.
Lastly, the well-known $\calO(\alpha\Gamma/M)$ suppression of nonfactorizable 
corrections for inclusive cross sections \cite{nonfac} could be probed for more 
complex scale hierarchies.

\acknowledgments %===================================================================
We would like to thank S.~Dittmaier for useful comments after carefully reviewing 
the manuscript, as well as D.~Berdine and D.~Rainwater for numerous useful 
discussions.  This work was supported by the BMBF, Germany (contract 05HT1WWA2).


\begin{thebibliography}{99}
\bibitem{footnote1}
For unstable particles with spin, $\sigma_\text{NWA} = \sigma_p\cdot\text{BR}$ + corrections if spin correlations are not neglected as seen explicitly after a Fierz transformation.
\bibitem{footnote2}
$t$ is defined with the two uppermost external momenta in all graphs.
\bibitem{footnote3}
In general, a suitable gauge has to be chosen and applied throughout.
\end{thebibliography}
\end{document}